\documentclass[useAMS,usenatbib]{mn2e}
\usepackage{epsf}
\usepackage{amssymb}
\usepackage{graphicx}
\usepackage[usenames]{color}
\usepackage{fmtcount}

\newcommand {\bc}{\begin {center}}
\newcommand {\ec}{\end {center}}\newcommand {\be}{\begin {equation}}
\newcommand {\ee}{\end {equation}}
\newcommand {\disp}{\displaystyle}
\title[Gas perturbations in cool cores of galaxy clusters]{Gas perturbations in cool cores of galaxy clusters: effective equation of state, velocity power spectra and turbulent heating}
\author[Zhuravleva et al.]{I. Zhuravleva$^{1,2}$\thanks{zhur@stanford.edu}, S. W. Allen$^{1,2,3}$, A. B. Mantz$^{1,2}$, N. Werner$^{4,5,6}$\\
$^1$Kavli Institute for Particle Astrophysics and Cosmology, Stanford University, 452 Lomita Mall, Stanford, California 94305-4085, USA\\
$^2$Department of Physics, Stanford University, 382 Via Pueblo Mall,
Stanford, California 94305-4060, USA\\
$^3$SLAC National Accelerator Laboratory, 2575 Sand Hill Road, Menlo Park, CA 94025, USA\\
$^4$MTA-E$\ddot{o}$tv$\ddot{o}$s University Lend$\ddot{u}$let Hot Universe Research Group, P$\acute{a}$zm$\acute{a}$ny P$\acute{e}$ter s$\acute{e}$t$\acute{a}$ny 1/A, Budapest, 1117, Hungary\\
$^5$Department of Theoretical Physics and Astrophysics, Faculty of Science, Masaryk University, Kotl$\acute{a}$$\check{r}$sk$\acute{a}$ 2, Brno, 611 37, Czech Republic\\
$^6$School of Science, Hiroshima University, 1-3-1 Kagamiyama, Higashi-Hiroshima 739-8526, Japan
}

\begin{document}

\date{Accepted .... Received ...}

\pagerange{\pageref{firstpage}--\pageref{lastpage}} \pubyear{2017}

\maketitle

\label{firstpage}

\begin{abstract}
We present the statistical analysis of X-ray surface brightness and gas density fluctuations in cool cores of ten, nearby and bright galaxy clusters that have deep {\it Chandra} observations and show observational indications of radio-mechanical AGN feedback. Within the central parts of cool cores the total variance of fluctuations is dominated by isobaric and/or isothermal fluctuations on spatial scales $\sim$ 10-60 kpc, which are likely associated with slow gas motions and bubbles of relativistic plasma. Adiabatic fluctuations associated with weak shocks constitute less than 10 per cent of the total variance in all clusters. The typical amplitude of density fluctuations is small, $\sim$ 10 per cent or less on scales of $\sim$ 10-15 kpc. Subdominant contribution of adiabatic fluctuations and small amplitude of density fluctuations support a model of gentle AGN feedback as opposed to periodically explosive scenarios which are implemented in some numerical simulations. Measured one-component velocities of gas motions are typically below 100-150 km/s on scales $<$ 50 kpc, and can be up to $\sim$ 300 km/s on $\sim$ 100 kpc scales. The non-thermal energy is $<$ 12 per cent of the thermal energy. Regardless of the source that drives these motions the dissipation of the energy in such motions provides heat that is sufficient to balance radiative cooling on average, albeit the uncertainties are large. Presented results here support previous conclusions based on the analysis of the Virgo and Perseus Clusters, and agree with the {\it Hitomi} measurements. With next generation observatories like {\it Athena} and {\it Lynx}, these techniques will be yet more powerful.
\end{abstract}


\section{Introduction}
Hot gas ($T\sim 10^7$ -- $10^8$ K) in intracluster medium (ICM) is perturbed by many physical processes. On large spatial scales, 100s of kpc, perturbations are mostly produced by mergers, while various plasma instabilities disturb gas on sub-kpc scales. On intermediate scales the gas is disturbed by motions of galaxies, Active Galactic Nuclei (AGN) feedback, gas cooling, etc. Therefore, by probing gas perturbations on different spatial scales, a variety of fundamental physical processes in the ICM can be studied.

Usually, volume-filling perturbations are analyzed statistically by measuring power spectra of fluctuations present in the maps of the X-ray surface brightness, temperature, pressure, Sunyaev-Zeldovich effect, or, if scale information is not important, by looking at deviations from the average thermodynamic profiles. In many works \citep[e.g.][]{Shi01,Sch04,Fin05,Kaw08,Gu09,Chu12,Hof16,Kha16,Wer16,Eck17} mostly large-scale substructure originating from different stages of cluster mergers is probed. Here, instead, we are interested in understanding the physics of perturbations on small spatial scales in the bulk of the gas within cool cores of galaxy clusters that show observational indications of radio-mechanical AGN feedback. 

In the standard picture of radio-mechanical AGN feedback, the energy from a central supermassive black hole is released in a form of bubbles of relativistic plasma, which appear as cavities in X-ray images \citep[e.g.][see also review by Fabian 2012]{Chu00,McN00}. Initial rapid inflation of bubbles produces weak shocks with typical Mach numbers of $\sim 1.1$ --$1.5$ \citep[e.g.][]{For07,Gra08,Nul13b} that propagate quickly through cluster atmosphere, heating the gas\footnote {Since shocks are weak, $M^2-1 \ll1$, the amount of heat injected by each such shock into ICM is small \citep{Lan59}.}. Numerical simulations of shock propagation through cluster gas show that, in order to reproduce the observed properties of shocks, and the sizes of shock-heated gas regions and bubbles, the AGN should inflate bubbles slowly, transporting most of its energy to the enthalpy of bubbles, and only $\sim 20$ per cent of the injected energy should be carried by the shocks \citep{For17,Zhu16,Tan17}. This picture of a long-duration outburst is consistent with previous findings that AGN bubbles are powerful enough to balance radiative cooling in cluster cores \citep[e.g.][]{Bir04,Raf06,Hla12}. Once the bubbles are inflated, they rise buoyantly in the cluster atmosphere with the velocity that is smaller than the sound speed \citep{Chu01,Rey05}. During the rise, bubbles perturb the ICM and deposit their energy upward. The dynamics of the bubbles and their energy deposition strongly depends of the motions of the ambient gas and ICM inhomogeneities \citep[e.g.][]{Hei06}. The details of the heating transport processes are still a subject of much debate. One possibility is that the energy from the rising bubbles is transported to the ICM through the in situ generated turbulence in wakes and other slow motions of gas, which in stratified cluster atmosphere are essentially internal gravity waves \citep[e.g.][]{Chu02,Omm04}. The internal waves efficiently spread the energy in the perpendicular to radial direction, and are trapped within the core region since their buoyancy frequency is decreasing with radius \citep[e.g.][]{Bal90}. Therefore, these waves inevitably interact with each other, becoming non-linear and eventually dissipating in the gas. Other actively discussed possibilities of the energy transport from bubbles to the ICM include streaming and diffusion of cosmic ray protons \citep[e.g.][]{Loe91,Guo08,Pfr13,Jac17}, and mixing of the hot bubble gas \citep[e.g.][]{Hil16,Hil17}. However, if there are weak magnetic fields in the ICM, the drapping of magnetic field lines may affect the evolution of bubbles \citep{Dur08} and the efficiency of their mixing \citep{Wei17}. More explosive mechanisms of AGN feedback, in which bubbles are inflated rapidly and the majority of AGN-injected energy is transported into compressive modes (shocks and sound waves) are also studied in details \citep[e.g.][]{Rey15,Yan16}.

In order to better understand the complex physics of AGN-ICM interaction, we perform a statistical analysis of gas perturbations in cluster cores. It has been shown that the inner region on the Perseus Cluster core (radii $\lesssim 100$ kpc) is predominantly disturbed by the activity of AGN, and the outer region within the core by merger events \citep[][]{Fab11}. Therefore, we primarily focus on fluctuations in the inner cool-core regions, defined as radii less than half the cooling radius (see Section \ref{sec:data_analysis}).  Also, we restrict the analysis to spatial scales that are smaller than any characteristic scale present in a system (e.g. the size of the cool core, $\sim 100$ kpc; the entropy/pressure/density scale height, $\sim 100$ kpc; the size of rising bubbles, $\sim$ tens of kpc). Fluctuations on such spatial scales have been analyzed in few cool core clusters so far: AWM7 \citep{San12}, Perseus and Virgo \citep[e.g.][]{Zhu14b, Are16}, and Centaurus \citep{Wal15}.  Clearly, these measurements are challenging as the signal from fluctuations on such small scales is often dominated by Poisson noise, disturbed by the point spread function (PSF) of X-ray instruments, and affected by unresolved point sources  and projection effects \citep[e.g.][]{Sch04,Chu12}. 

The {\it Chandra} X-ray observatory currently provides the best data for such analysis because of its superb spatial resolution. Based on the analysis of fluctuations present in exceptionally deep {\it Chandra} images of the cores of the Perseus and Virgo clusters, it was shown that most energy in perturbations on spatial scales $\sim 20$ kpc and less is associated with isobaric (e.g. slow gas displacements) and a combination of isothermal (e.g. bubbles) and isobaric fluctuations, respectively \citep[][hereafter A16 and Z16, respectively]{Are16,Zhu16}. Under some assumptions, the velocity power spectra can be derived from the measured density power spectra \citep[e.g.][]{Zhu15}. Remarkably, direct velocity measurements provided by the {\it Hitomi} satellite in the innermost $\sim 60$ kpc region of the Perseus Cluster are consistent with the indirect measurements through density fluctuations, within the uncertainties \citep{Hit16}. Knowing the velocity amplitude and associated with it spatial scale, the turbulent heating rate can be calculated. It was shown that the dissipation of gas motions in cluster cores provides enough energy to offset radiative cooling \citep[Zhuravleva et al. 2014, see also theoretical studies by][]{Fuj04,Ban14}. Note that even if it is likely that the activity of the central AGN in Perseus and Virgo triggers most of the measured gas motions, there could be additional contribution from turbulent motions generated when jet-driven shocks interact with pre-existing bubbles \citep{Fri12}, by motions of galaxies \citep{Bal90,Gu13}, and by mergers with subhalos and galaxies \citep[e.g.][]{Nor99,Dol05,Vaz11,Min15,Lau17,Bou17}. 

Here, we expand the fluctuations analysis to a sample of bright, nearby cool-core clusters. We check whether the main conclusions about the effective equation of state of gas perturbations and the role of turbulence in heating processes from the fluctuation analysis in Perseus and Virgo are valid in other bright cool cores. The agreement with the {\it Hitomi} results additionally motivates us to expand the velocity measurements in these clusters.

The structure of the paper as follows. Our sample of cool core clusters in described in Section 2. We discuss the details of the data analysis and our methods in Sections 3 and 4, respectively. Section 5 shows the results, which are then discussed in Section 6. Finally, in Section 7, we summarize the main results and conclusions.

\section {Sample selection}
\label{sec:sample}

\begin{figure}
\includegraphics[trim=20 200 20 90,width=0.49\textwidth]{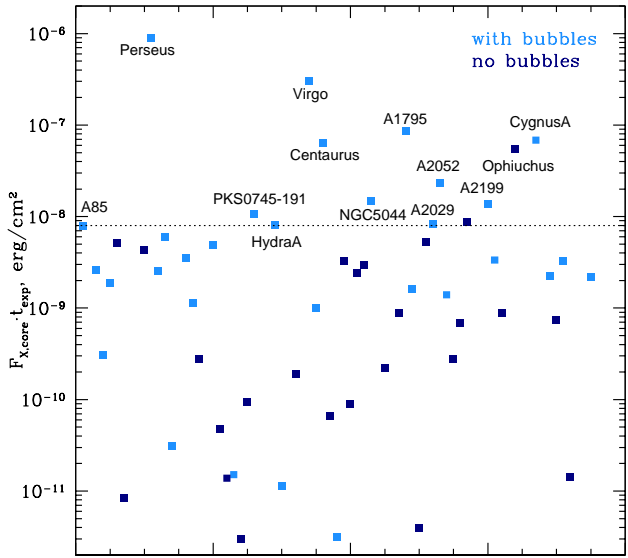}
\caption{Selection of galaxy clusters for our analysis: comparison of the product of X-ray flux calculated within a cool core times total exposure time of {\it Chandra} observations available in archive for all clusters in the complete sample of the brightest clusters \citep{Bir12}. {\it Chandra}  observations with large offset are excluded. Light blue points: clusters with detected bubbles of relativistic plasma; dark blue points: clusters without bubbles. Dotted line determines the threshold for the cluster selection for our sample (see Section \ref{sec:sample}).   
\label{fig:sample}
}
\end{figure}

\begin{table*}
\centering
\caption{Basic properties of galaxy clusters in our sample: cluster names, {\it Chandra} ObsIDs used for the anlaysis, total exposures in ks, redshifts from the NED  database (https://ned.ipac.caltech.edu), total Galactic HI column densities \citep{Kal05}, half-radii of cool cores in kpc and arcmin (see Section 3), energy bands of images used for the fluctuation analysis.}
\begin{tabular}{@{}llccccc@{}}
\hline
Name  & ObsIDs & Exposure, & Redshift & N$_{\rm H}$,     &  $r_{\rm cool}$/2, &  Energy bands,\\
           &              & ks                     &               & 10$^{20}$ cm$^{-2}$   &  kpc / arcmin  &       keV\\
\hline
Perseus     & 3209, 4289, 4946--4953, 6139,   & 1474  & 0.0179 & 13.6 & 87 / 4.1 & soft: 0.5--3.5\\
                  & 6145, 6146, 11713--11716,          &           &             &                                       &              & hard: 3.5--7.5\\
                  & 12025, 12033, 12036, 12037        &           &             &                                       &              &                      \\ 
Virgo          & 2707, 3717, 5826--5828,             & 636     &  0.0036& 2.38 & 42 / 9    & soft: 1.0--3.5 \\
                  & 6186, 7210--7212, 11783             &            &             &                                      &              & hard: 3.5--7.5\\
Centaurus  & 4190, 4191, 4954, 4955,              & 747     &  0.0114& 8.56   & 50 / 3.7 & soft: 1.0--3.5 \\
                  & 5310, 16223--16225, 16534,       &           &              &                                       &              & hard: 3.5--7.5\\
                  & 16607--16610                             &            &              &                                       &              &                     \\
A2052       & 890, 5807, 10477--10480,           & 663     & 0.0355 & 2.71 & 64 / 1.6 & soft: 0.5--3.5\\
                 & 10879, 10914--10917                  &            &              &                                       &               & hard: 3.5--7.5 \\
A2199       & 10803--10805, 10748                  & 121     & 0.0303 & 0.89 & 69 / 1.9 & soft: 0.5--3.5\\
                 &                                                       &            &              &                                      &              & hard: 3.5--7.5\\
A1795       & 3666, 5286--5290, 6159--6163,  & 1400   & 0.0625 & 1.19 & 86 / 1.2  & soft: 0.5--3.5\\
                 & 10898--10901, 12026--12029,    &            &               &                                     &               & hard: 3.5--7.5\\
                 & 13106--13113, 13412--13417,    &            &               &                                     &               &                         \\
                 & 14268--14275, 15485, 15486,     &            &               &                                     &               &                         \\      
                 & 15488--15492, 16432--16439,    &            &               &                                     &               &                         \\      
                 & 16465--16472, 17397--17411,    &            &               &                                     &               &                         \\      
                 & 17683--17686, 18423--18439     &            &               &                                     &               &                         \\      
PKS0745-191 & 2427, 6103, 7694, 12881       & 153     & 0.1028   & 41.8 & 105 / 0.09 & soft: 1.5--3.5\\
                  &                                                     &            &               &                                     &               & hard: 3.5--7.5\\
A85           & 15173, 15174, 16263, 16264       & 159     & 0.0551 & 2.78 & 76 / 1.2  & soft: 0.5--3.5\\
                  &                                                     &           &               &                                     &               & hard: 3.5--7.5\\
Hydra A     & 4969, 4970                                  & 198    & 0.0549 & 4.68 & 95 / 1.5 & soft: 0.5--3.5 \\
                  &                                                     &           &               &                                     &               & hard: 3.5--7.5\\
A2029       & 891, 6101, 4977                          & 109     & 0.0773 & 3.25 & 92 / 1.1 & soft: 0.5--3.5\\
                 &                                                     &           &               &                                     &               & hard: 3.5--7.5\\
\hline
\label{tab:sample}
\end{tabular}
\end{table*}

In this work, we present an analysis of a subsample of galaxy clusters drawn from the \citet[][]{Bir12} complete sample of the brightest galaxy clusters. In order to probe fluctuations on the spatial scales of interest (few tens of kpc or smaller), we select clusters with the most photon counts in their cores. Fig. \ref{fig:sample} shows the product of the X-ray flux from the cluster core, $F_{\rm X,core}$, and the total exposure time $t_{\rm exp}$ of the respective {\it Chandra} observations, which should be proportional to the number of recorded source counts. The Perseus Cluster is clearly an outlier due to its high X-ray brightness and exceptionally long, $\sim 1.4$ Ms, {\it Chandra} observations. The Virgo and Centaurus Clusters are another examples of bright clusters with deep ($> 600$ ks) observations. The amplitude of gas emissivity fluctuations has been reliably measured in these clusters on spatial scales down to $\sim 5-30$ kpc \citep[e.g.][]{Zhu15,Wal15,Are16}.  The next most promising targets for fluctuation analysis are A1795, A2052, and A2199. Our analysis, described below, shows that high Poisson noise in A85 and A2029 limits the measurements on spatial scales smaller than $\sim 20$ -- $30$ kpc. Clusters with even lower values of  $F_{\rm X,core}\cdot t_{\rm exp}$ are, clearly, less promising.

Our final sample includes $10$ galaxy clusters listed in Table \ref{tab:sample}. Note that several objects with large $F_{\rm X,core}\cdot t_{\rm exp}$ are not included in our sample: namely, Cygnus A, which is a FRII type object \citep[e.g.][]{You02,Cho12} and the Ophiuchus Cluster with dynamically disturbed unusually small ($\sim 30$ kpc only) cool core that does not show any hints of the radio-mechanical feedback \citep[e.g.][]{Mil10,Wer16}. For the homogeneity of our sample and analysis procedure we also do not consider NGC5044, a group of galaxies filled with relatively cool gas, with temperatures of $\sim 1.5$  keV and below \citep[e.g.][]{Buo03}.

\section {Data analysis and X-ray images}
\label{sec:data_analysis}
\begin{figure*}
\includegraphics[trim=0 0 0 0,width=\textwidth]{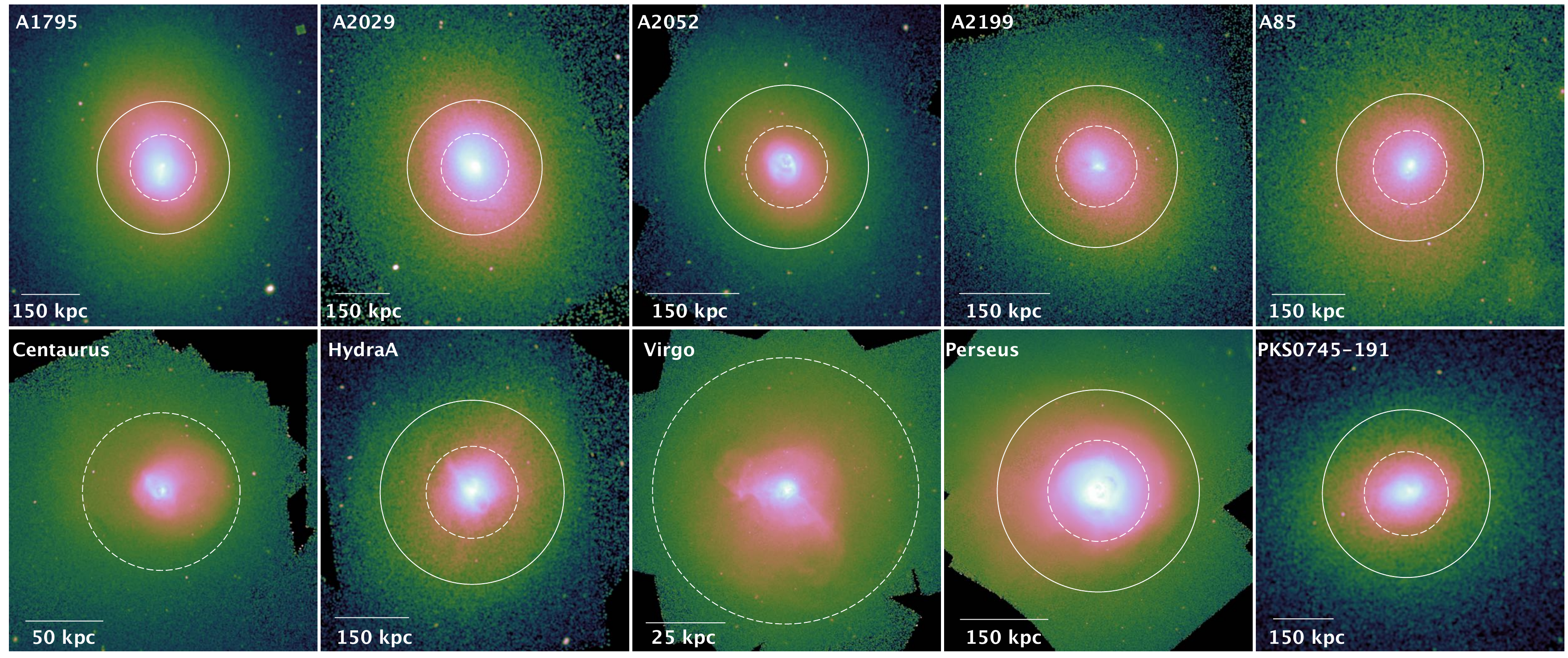}
\caption{Mosaic {\it Chandra} images of galaxy clusters in our sample. Cool core regions are indicated with the solid circles. Dashed circles show the inner regions with the half cool-core radius. For display purposes, all images are lightly smoothed with a 2 arcsec Gaussian.   
\label{fig:images}
}
\end{figure*}

For all clusters in our sample we use public {\it Chandra} data available in archive. ObsIDs are summarized in Table \ref{tab:sample}. The data are reprocessed using standard algorithms developed by \citet{Vik05}, applying the latest calibration data. Correcting for the exposure and vignetting effects, and subtracting the background, a mosaic image of each cluster is produced (see Fig. \ref{fig:images}).   

Point source candidates are identified using the {\scriptsize WVDECOMP} tool. The significance of each point source detection is verified using procedure described in \citet[][]{Zhu15}. Accounting for the PSF shape in combined images, the verified point sources are excised from all images of clusters.

To determine thermodynamic properties of hot gas in all clusters, we extract a set of azimuthally averaged,  projected spectra from concentric annuli. These projected spectra are then deprojected using the procedure described in \citet{Chu03}. The resulting spectra for every shell are fitted in the $0.6-8.5$ keV band\footnote{We ignore photons with the energies $< 1.5$ keV in the PKS0745-191 cluster since it lies close to the plane of our Galaxy. The line-of-sight absorption column density is high, $n_H=4.18\cdot 10^{21}$ cm$^{-2}$, see Table \ref{tab:sample}.}  using {\scriptsize XSPEC}  and a single-temperature {\scriptsize APEC} plasma model \citep{Smi01,For12} based on ATOMDB version 3.0.7. The abundance of heavy elements is normally fixed to 0.5 with respect to the solar abundances of \citet{And89}. Only in Hydra   A and Virgo the abundance is treated as a free parameter in the models, as doing so slightly changes the deprojected temperature. 

Using deprojected radial profiles of the electron number density and temperature, we calculate the gas cooling time as
\be
t_{\rm cool}=\frac{3}{2}\frac{(n_{\rm e}+n_{\rm i})k_{\rm B}T}{n_{\rm e}n_{\rm i}\Lambda(T)},
\ee
where $n_{\rm e}$ and $n_{\rm i}$ are the number densities of electrons and ions, respectively, $k_{\rm B}$ is the Boltzmann constant, $T$ is the gas temperature and $\Lambda(T)$ is the normalized cooling function for $0.5$ Solar metallicity tabulated by \citet{Sut93}. For each cluster we define a cooling radius, $r_{\rm cool}$, as the distance from the cluster center where the cooling time, $t_{\rm cool}$, is equal to the Hubble time; see Table \ref{tab:sample} for the summary.  

In order to probe the effective equation of state of gas perturbations (see Sections 4 and 5.1), we prepare X-ray images of galaxy clusters in two different energy bands. For the majority of objects, the $0.5-3.5$ keV band is used to define the ``density'' band. The X-ray emissivity in this band is independent of gas temperature for clusters with a mean temperature larger than $\sim 3$ keV. The emissivity in the hard, $3.5-7.5$ keV, band is ``temperature-dependent". We refer the reader to Z16 for the details about the choice of energy bands. Note that the Virgo and Centaurus clusters are the coolest objects in our sample. Their gas temperatures drop down to $\sim 1$ keV in the innermost regions. Therefore, for these clusters the ``density" band is different, $1.0-3. 5$ keV \citep[][A16]{For07}. Lastly, the amplitude of fluctuations in the $0.5-3.5$ keV image of PKS0745-191 can be affected by the high Galactic column density towards this cluster. In order to minimize the effect, we exclude photons with the energies lower than 1.5 keV from the soft-band image. 

\begin{figure*}
\includegraphics[trim=0 0 0 0,width=\textwidth]{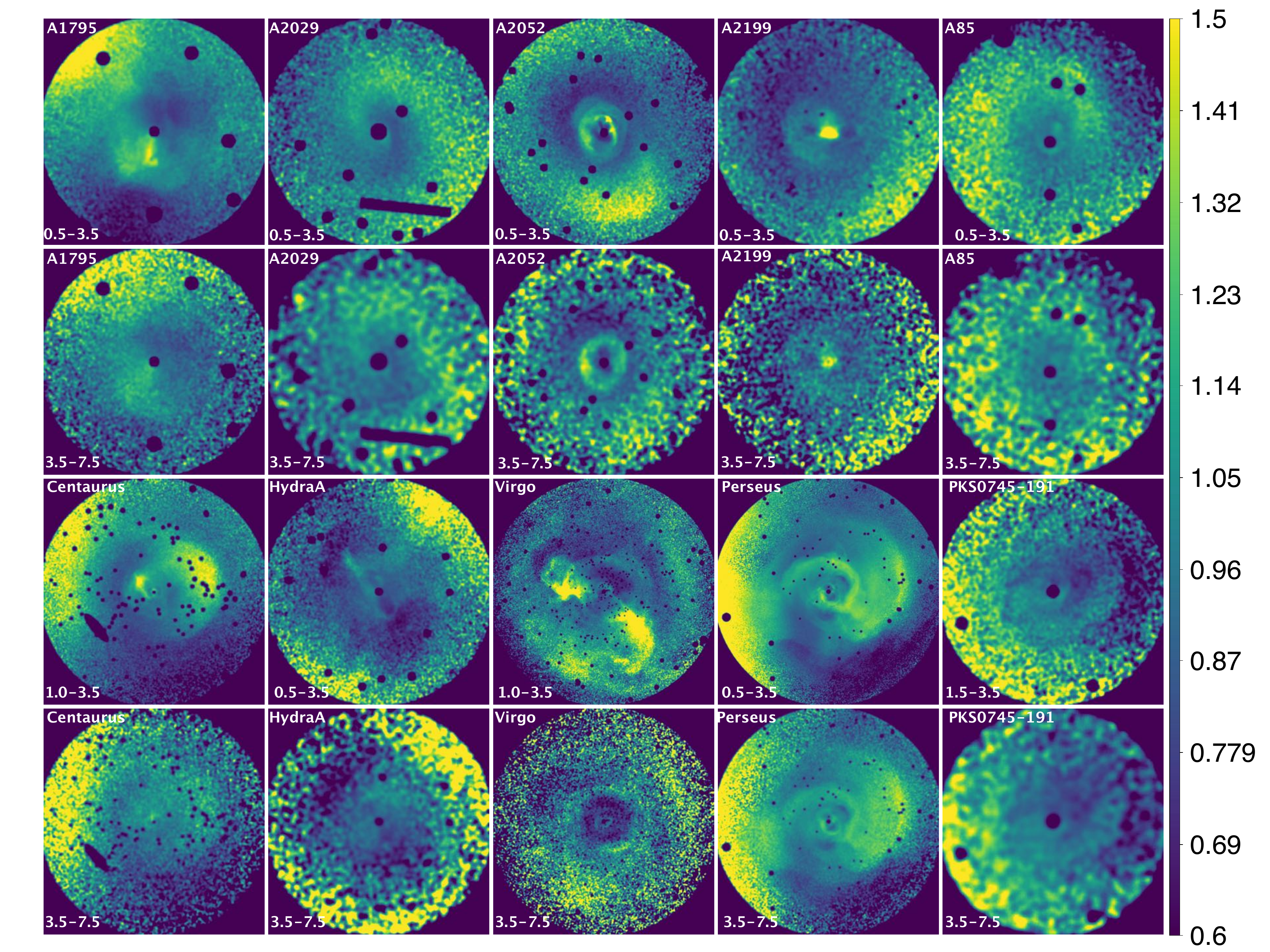}
\caption{Residual X-ray images of galaxy clusters in our sample in the soft (1st and 3rd rows) and hard (2nd and 4th rows) bands. Only cool core regions are shown. Point sources, central AGNs and projected edge-on galaxies are excised. The spatial scales and color scales of paired images are the same. All images are lightly smoothed for visual purposes.   
\label{fig:images_resid}
}
\end{figure*}

For the majority of objects in our sample the {\it Chandra} PSF is smallest within the inner, half cool-core regions. Also, Poisson noise is minimized in these regions. Therefore, we choose to probe fluctuations in two independent regions within the cool core instead of the whole core: the innermost region within half of the cooling radius, and the outer $r_{\rm cool}/2 < r < r_{\rm cool}$ annulus. Of course, for some clusters, the analysis can be done in finer annuli \citep[e.g.][]{Zhu14b}. However, in order to perform homogeneous analysis for all clusters, we consider only these two regions in this work. For the two nearest clusters, Virgo and Centaurus, the {\it Chandra} data covers only the innermost regions. Solid and dashed circles in Fig. \ref{fig:images} indicate the chosen regions. 

For the analysis of gas perturbations, the underlying large-scale surface brightness gradient is removed by fitting an elliptical $\beta-$model to the images in both bands independently and dividing the images by the best-fitting models. For the Perseus, Virgo, Hydra A and Centaurus clusters, spherical $\beta-$models are used instead, since the elliptical models within the considered regions do not change the results. The residual images of cool-core regions in both energy bands are shown in Fig. \ref{fig:images_resid}. Note that the choice of the  ``unperturbed" models of clusters does not affect our results, as long as the models are smooth, since we are considering fluctuations on relatively small spatial scales.

Looking at residual images, one can notice a lot of substructures. Many of the prominent features have been studied earlier. In all clusters there are multiple cavities filled with relativistic particles \citep[e.g.][]{Bir12}, which indicate the activity of central AGN. In some clusters, weak shocks are found (A2052: Blanton et al. 2011; A2199: Nulsen et al. 2013; Centaurus: Sanders et al. 2016; Hydra A: Gitti et al. 2011; Virgo: Forman et al. 2007; Perseus: Fabian et al. 2011). There are also spiral structures associated with sloshing of the gas, likely the result of mergers (A1795: Markevitch et al. 2001; A2029: Paterno-Mahler et al. 2013; A2052: Blanton et al. 2011; A2199: Nulsen et al. 2013, A85: Kempner et al. 2002; Centaurus: Sanders et al. 2016; PKS0749-191: Sanders et al. 2014). Similar structures can be formed as the result of AGN-ICM interaction \citep[see e.g. inner spiral structure in Perseus, ][]{Fab11}. It is likely that gas perturbations are mostly driven by AGN feedback processes in the inner regions in the Perseus, Virgo, Hydra A, Centaurus, and A2052 clusters. In the rest of clusters the contribution from mergers could be as significant. 

Some particularly bright features can dominate the signal in the measured power spectrum of fluctuations. We exclude them from the images used for the fluctuation analysis, since we are interested in perturbations in the bulk of the gas, which are often weak. Namely, we do not include inner 30 kpc region of enhanced pressure due to shocks in A2052 \citep[][]{Bla11}, the inner 14 kpc in A2199 \citep[very steep surface brightness, see e.g.][]{Joh02,San06}; cold outflows in Virgo \citep{For07}; the inner 30 kpc in Perseus, with very sharp edges associated with inner bubbles and their surrounding shocks \citep{Fab11}; and the high-metallicity region in Centaurus \citep[asymmetric, $\sim 30-40$ kpc, see ][]{All94,San16}.

\section{Power spectra analysis}

The power spectra of surface brightness fluctuations present in the X-ray images are measured using the modified $\Delta-$variance method \citep{Oss08,Are12}, which is specifically designed to calculate a low-resolution power spectrum from non-periodic data with gaps. Subtracting Poisson noise and deprojecting the measured power spectra, accounting for the global geometry of the cluster, the 3D power spectrum of volume emissivity fluctuations in both bands, $P_{k,aa}$ and $P_{k,bb}$, and the cross-spectrum of fluctuations in two energy bands, $P_{k,ab}$, are calculated\footnote{The deprojection procedure is valid only if the amplitude of emissivity fluctuations is small, i.e. $\delta n_k/n \ll 1$.}. The deprojected spectra are then corrected for the {\it Chandra} PSF and contribution of unresolved point sources; the latter is insignificant in all the cool cores studied here. The details of each step of the analysis are described in \citet{Chu12,Zhu15}. Following those works, instead of power spectra, we express our measurements as  the characteristic amplitude $A_{k,ab}=\sqrt{4\pi P_{k,ab}k^3}$ (similarly for $A_{k,aa}$ and $A_{k,bb}$), which is a proxy for the root-mean-square of emissivity fluctuations at a given wavenumber, $k=1/l$. Note that, since the temperature dependence of X-ray emissivity is weak in the soft band, the amplitude of emissivity fluctuations, $A_{k,aa}$, in that band is equivalent to the amplitude of gas density fluctuations multiplied by 2.  

Our first goal is to understand the nature of gas perturbations, i.e. to probe their apparent (or ``effective") equation of state, which measures the correlation between gas density and temperature fluctuations. Following A16 and Z16, we consider only three types of perturbations: (i) isobaric, i.e. any local changes of gas entropy in pressure equilibrium (e.g. any slow gas motions, including subsonic turbulence, gas sloshing, or, more generally, linear and non-linear internal waves); (ii) adiabatic perturbations associated with weak shocks (sound waves) with $M-1\ll1$, which do not change gas entropy substantially; and (iii) isothermal, which could be associated with the bubbles of relativistic plasma that appear as cavities in the X-ray images, i.e. a drop of gas density without apparent changes in gas temperature. For pure isobaric, adiabatic and isothermal perturbations, the proportionality coefficients between temperature and density perturbations are $-1$, $2/3$ and $0$, respectively. If the typical amplitude of fluctuations is small, one can find the response of the X-ray emissivity to these three types of fluctuations; namely, if the dominant type of fluctuations in gas with $T > 3$ keV is isobaric (adiabatic) in nature, the amplitude of emissivity fluctuations in the soft (hard) band will be larger than in the hard (soft) band. For isothermal fluctuations, the amplitudes in both bands are the same.  For given energy bands and gas temperatures, the ratio of volume emissivity perturbations for the different types of perturbations can be predicted (see the image arithmetic techniques by Churazov et al. 2016).

In the real ICM, all these processes are likely to be present at some level. For a given gas temperature, we can predict the coherence and the ratio of the amplitudes of fluctuations in the soft and hard bands for any mixture of isobaric, adiabatic and isothermal types of fluctuations, assuming that they are uncorrelated. If $\alpha_i^2=S_{k,i}/\sum\limits_j S_{k,j}$, where $S_{k,i}$ is the power spectrum of density fluctuations of the {\it i}th type, $\alpha_{\rm isob.}^2+\alpha_{\rm adiab.}^2+\alpha_{\rm isoth.}^2=1$. We calculate the maps of the expected values of coherence and ratio as a function of $\alpha_{\rm isob.}$ and $\alpha_{\rm adiab.}$, setting $\alpha_{\rm isoth.}=\sqrt{1-\alpha_{\rm isob.}^2-\alpha_{\rm adiab.}^2}$ for any combination of the three types of perturbations. These predictions are then used for the interpretation of the observed values of coherence and ratio to find the relative contribution of each type of perturbations to the observed total variance of fluctuations (for the details, see A16, Z16). 

From observations, we measure scale-dependent coherence $C(k)$ and ratio $R(k)$ of fluctuations through the measured power spectra of gas fluctuations in two energy bands and their cross-spectrum, namely:
\be
C(k)=\frac{P_{k,ab}}{\sqrt{P_{k,aa}P_{k,bb}}}
\ee  
and
\be
R(k)=\frac{P_{k,ab}}{P_{k,aa}}=C(k)\sqrt{\frac{P_{k,bb}}{P_{k,aa}}}
\ee
respectively.

Velocity power spectra are measured from the power spectra of density fluctuations (i.e. spectra of emissivity fluctuations in the soft X-ray images, $A_{k,aa}$, divided by 2). Simple theoretical arguments \citep{Zhu14a} show that, in stratified cluster atmospheres, the amplitude of density fluctuations, $\delta n_{\rm k}/n$, and one-component velocity, $V_{\rm 1,k}$, are proportional to each other at each wavenumber $k$ within the inertial range of scales,
\be
\frac{\delta n_{\rm k}}{n}=\eta \frac{V_{\rm 1,k}}{c_{\rm s}},
\label{rel:rhov}
\ee
where $c_{\rm s}=\sqrt{\gamma\disp\frac{k_{\rm B} T}{\mu m_{\rm p}}}$ is the sound speed, $\gamma=5/3$ is the adiabatic index, $\mu=0.61$ is the mean particle weight in units of the proton mass, $m_{\rm p}$ is the proton mass. Cosmological numerical simulations of relaxed clusters suggest that the proportionality coefficient is $\eta \sim 1\pm 0.3$ \citep{Zhu14a}.  Hydrodynamic simulations of gas in static cluster atmospheres confirm this result and investigate how electron conduction and gas viscosity affect it \citep{Gas14}.

\begin{figure*}
\includegraphics[trim=0 0 0 0,width=\textwidth]{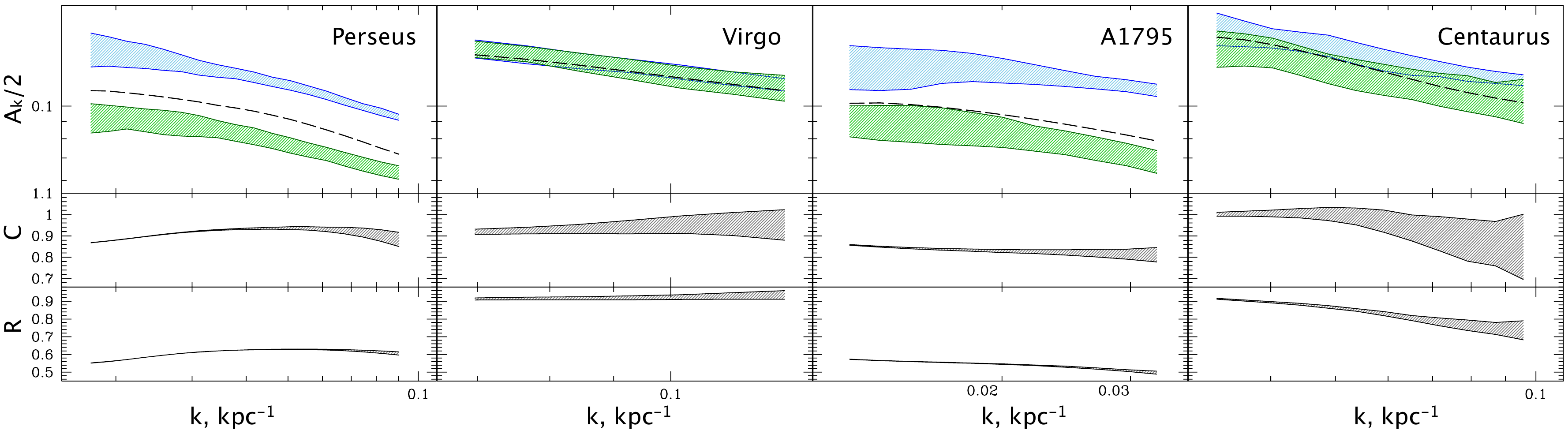}
\caption{Results of the cross-spectrum analysis of fluctuations in the inner parts of cool cores in the Perseus, Virgo, A1795 and Centaurus clusters. Top panels: the amplitude of the volume emissivity fluctuations divided by a factor of 2, which in the soft band corresponds to the amplitude of density fluctuations. Blue$/$Green: the amplitude measured from the soft/hard band images. The width of the regions reflects the $1\sigma$ statistical and stochastic uncertainties. Black dashed curves: the cross-amplitude of fluctuations. For visual clarity, we do not show the uncertainties, but take them into account when calculating $C$ and $R$. Only scales where the amplitude of fluctuations in both bands is least affected by systematic uncertainties and Poisson noise are shown. Middle and bottom panels: coherence, $C$, and ratio, $R$, obtained from the observed power spectra; see relations (2) and (3). 
\label{fig:cc}
}
\end{figure*}

\begin{figure*}
\includegraphics[trim=0 0 0 0,width=\textwidth]{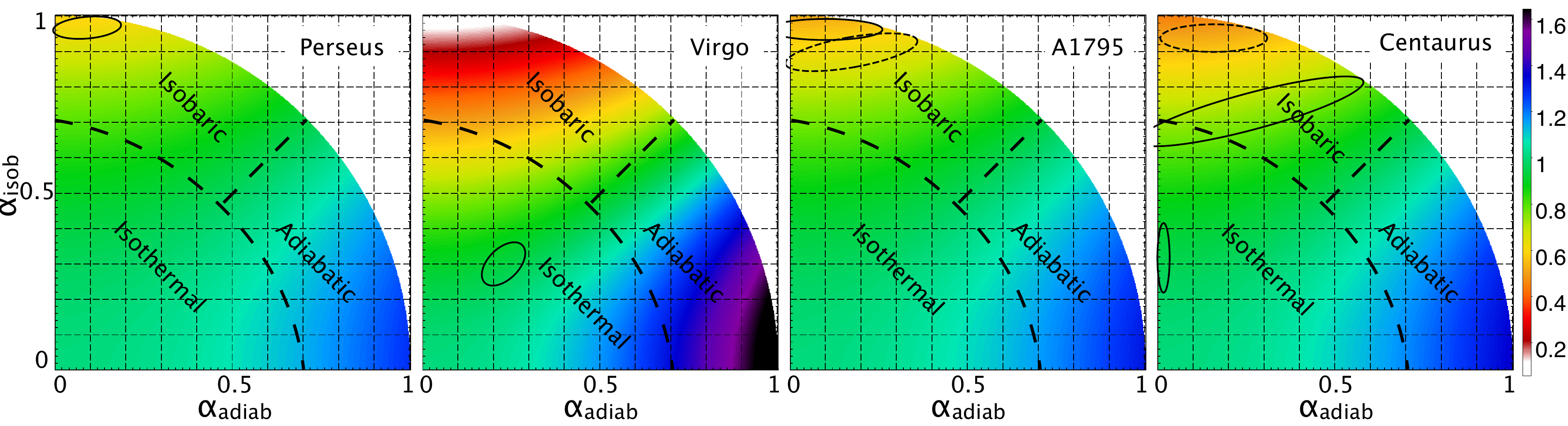}
\caption{Maps of fluctuation ratio ($R$) for a mixture of isobaric, adiabatic and isothermal perturbations for a subsample of clusters shown in Fig. \ref{fig:cc}. Color: the value of $R$; X-axis: contribution of adiabatic fluctuations $\alpha_{\rm adiab.}$; Y-axis: contribution of isobaric fluctuations $\alpha_{\rm isob.}$. The contribution of isothermal fluctuations is $\alpha_{\rm isoth.}=\sqrt{1-\alpha_{\rm adiab.}^2-\alpha_{\rm isob.}^2}$. The maps are schematically divided into three regions where one of the types of perturbations is dominant in terms of total variance. Black solid ellipses show the regions of $\alpha_{\rm adiab.}$, $\alpha_{\rm isob.}$ and $\alpha_{\rm isoth.}$ that correspond to the values of $C$ and $R$ shown in Fig. \ref{fig:cc} and Table 2. Dashed ellipse in A1795: same as solid ellipse, but obtained from the images excluding extended bright feature in the South shown in Fig. \ref{fig:images_resid} \citep{Fab01}. Dashed ellipse in Centaurus: results for the whole inner region, i.e. including the high-metallicity region \citep{San16}. Note that fluctuations are predominantly isobaric or isothermal.
\label{fig:regr}
}
\end{figure*}

\section{Results}
\label{sec:results}

 \subsection{The effective equation of state of gas perturbations}
\label{sec:eeos}
\begin{table*}
\centering
\caption{Summary of the results on the nature of gas perturbations in the inner half cool core region in all clusters from our sample: cluster name, spatial scales used for this part of the analysis, coherence and ratio measured on the smallest and the largest wavenumbers, mean temperature of gas used to generate $C$ and $R$ maps, approximate contribution of different types of fluctuations to the total variance. See Section \ref{sec:eeos} for details.}
\begin{tabular}{@{}lccccccl@{}}
\hline
Name & Scales, & $C(k_{\rm min})$ & $R(k_{\rm min})$ & $C(k_{\rm max})$ & $R(k_{\rm max})$& T, & Contribution to the total variance,\\
           & kpc                 &              &         &   &     &  keV               &  per cent                             \\
\hline
Perseus & $\sim 10 - 60$ &  0.8677$\pm$0.0004 & 0.5519$\pm$0.0003 & 0.88$\pm$0.03& 0.605$\pm$0.009 & 4.0 & i/b: $\sim$ 92; i/t: $\sim$ 7;  a/b: $<$ 1 \\
Virgo      & $\sim 8 - 15$   &  0.92$\pm$0.01    &  0.913$\pm$0.006 & 0.95$\pm$ 0.07& 0.94$\pm$0.02  & 1.8 & i/b: $\sim$ 9; i/t: $\sim$ 86; a/b: $\sim$ 5  \\
A1795    & $\sim 30 - 80$  & 0.858$\pm$0.001 & 0.5726$\pm$0.0005 & 0.812 $\pm$ 0.03 & 0.497 $\pm$0.008& 3.7 &  i/b: $\sim$ 90; i/t: $\sim$ 7;  a/b: $\sim$ 2 \\
Centaurus & $\sim 10 - 30$ & 1.001$\pm$0.009 & 0.914$\pm$0.003 &0.85$\pm$0.15 & 0.74$\pm$0.05& 3.0 & i/b: $\sim$ 12; i/t: $\sim$ 88; a/b: $<$ 1 ($k_{\rm min}$) \\
 &  & & & & & & i/b: $\sim$ 56; i/t: $\sim$ 35; a/b: $\sim$ 9 ($k_{\rm max}$) \\
A2029       & $\sim 60 - 90$ & 0.99$\pm$0.02 & 0.811$\pm$0.007 & 1.03$\pm$0.07 & 0.79$\pm$0.02& 6.2 & i/b: $\sim$ 76; i/t: $\sim$ 19;  a/b: $\sim$ 5  \\
A2052    & $\sim 24 - 35$ & 1.23$\pm$0.12 & 0.78$\pm$0.01 & 1.18$\pm$ 0.24&0.78$\pm$0.04 & 3.0 & i/b: $\sim$ 41; i/t: $\sim$ 50;  a/b: $\sim$ 9\\
A2199 & $\sim 30 - 33$ & 1.04$\pm$0.12 & 0.82$\pm$0.04 & 1.13$\pm$0.19& 0.81$\pm$0.05& 3.4 &  i/b: $\sim$ 41; i/t:  $\sim$ 56;  a/b: $\sim$ 3 \\
A85   & $\sim 49 - 87$ & 0.95$\pm$0.03 & 0.79$\pm$0.01 & 0.90$\pm$0.16& 0.70$\pm$0.05& 3.5 & i/b: $\sim$ 56; i/t: $\sim$ 38;  a/b: $\sim$ 6  \\
Hydra A & $\sim 56 - 75$ & 0.64$\pm$0.02 & 0.61$\pm$0.01 & 0.61$\pm$0.05& 0.54$\pm$0.01& 2.7 &  i/b: $\sim$ 71; i/t: $\sim$ 23;  a/b: $\sim$ 6\\
\hline
\label{tab:cc}
\end{tabular}
\end{table*}

\begin{figure}
\includegraphics[trim=0 0 0 0,width=0.4\textwidth]{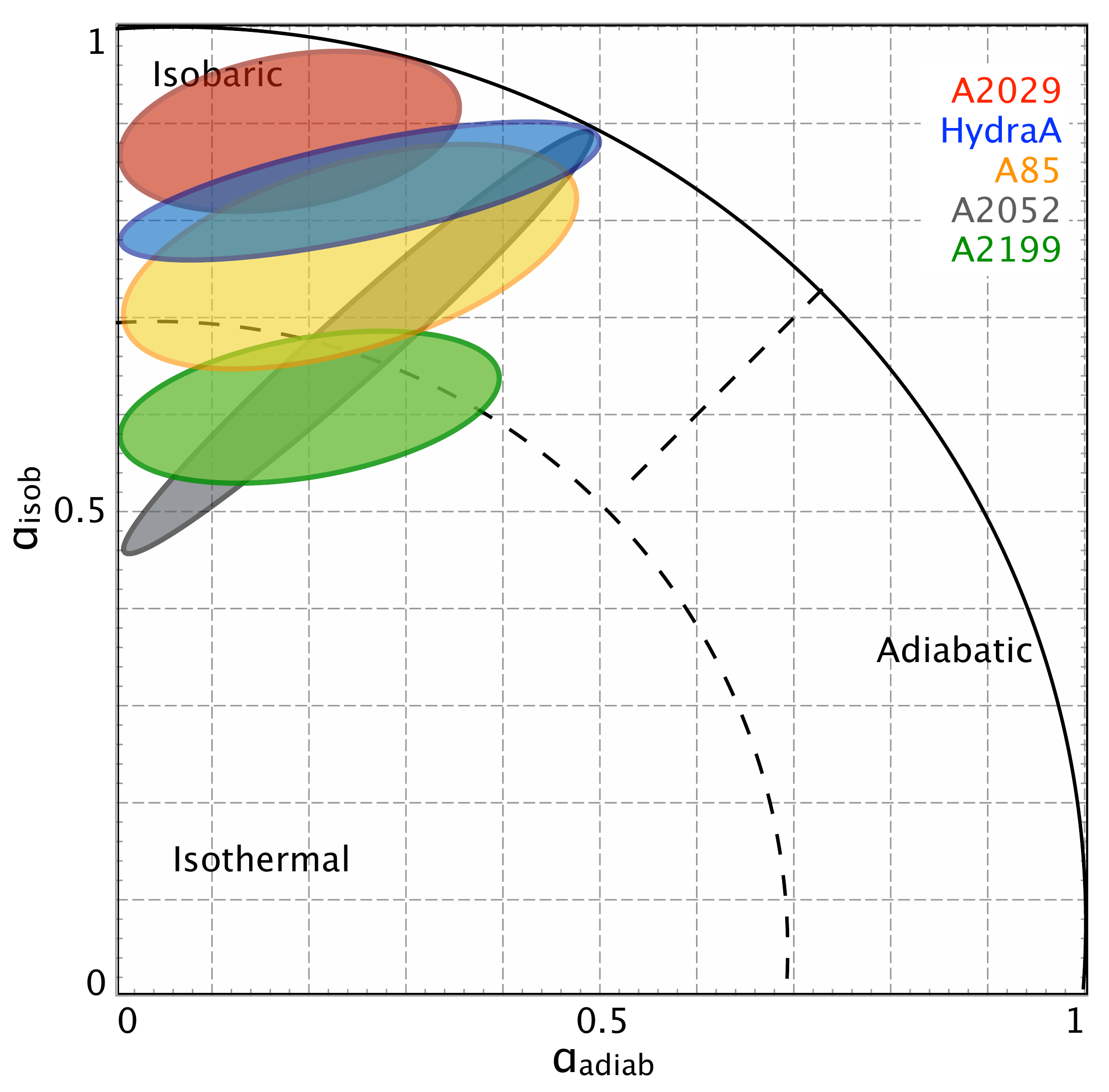}
\caption{Schematic map of $R$ and regions of $\alpha_{\rm adiab.}$, $\alpha_{\rm isob.}$ and $\alpha_{\rm isoth.}$ that correspond to the values of $C$ and $R$ for the rest of clusters in our sample. Note that fluctuations are predominantly isobaric or isobaric/isothermal.
\label{fig:reg_rest}
}
\end{figure}
We would first like to understand the nature of the dominant type of perturbations in the bulk of the gas and the relative contribution of each type (isobaric, adiabatic or isothermal) to the total variance of fluctuations.  As mentioned above, two regions in each cluster are considered: the inner half cool core (with excluded exceptionally bright features; see Section 3) and the outer part of the cool core (annulus). Fig. \ref{fig:cc} shows the results for the inner half-core regions in the Perseus, Virgo and Centaurus clusters. These clusters have the largest number of counts in the hard-band images, and the corresponding amplitudes of fluctuations are measured down to spatial scales of $\sim 10$ kpc or even lower in the case of Virgo. A1795 is also shown in Fig. \ref{fig:cc}, since fluctuations in the hard band are probed on a broad range of scales compared to other clusters. The amplitudes of the volume emissivity fluctuations are shown on spatial scales that are least affected by Poisson noise in the hard band and systematic uncertainties associated with the removal of global surface brightness profiles. In Perseus and A1795, the amplitude of fluctuations in the soft band is clearly larger than in the hard band on spatial scales of $\sim 10-60$ kpc and $\sim 30-80$ kpc, respectively, pointing to predominantly isobaric fluctuations. In the case of Virgo and Centaurus, the soft- and hard-band amplitudes are comparable, suggesting that fluctuations are mostly isothermal. 

Accounting for the cross-amplitude of emissivity fluctuations, the coherence and ratio are calculated (bottom panels in Fig. \ref{fig:cc}). For any combination of isobaric, adiabatic and isothermal fluctuations, the maps of $C$ and $R$ as a function of the $\alpha_{\rm isob.}$, $\alpha_{\rm adiab.}$ and $\alpha_{\rm isoth.}$ are generated, assuming the mean gas temperature (see Table 2) within the considered regions in each cluster. Finding the measured $C$ and $R$ on these maps, the relative contribution of each type of perturbations to the observed total variance of the fluctuations is obtained. Below, we use $C$ and $R$ measured on the smallest and the largest wavenumbers probed in each cluster. Fig. \ref{fig:regr} shows $R$ maps for the four clusters from Fig. \ref{fig:cc}. The loci of the measured $C$ and $R$ are shown with solid ellipses. Results for A1795 reveal that the mean value of $\alpha_{\rm adiab.}\simeq 0.15$, $\alpha_{\rm isob.}\simeq 0.95$ and $\alpha_{\rm isoth.}=\sqrt{1-\alpha_{\rm adiab.}^2-\alpha_{\rm isob.}^2}\simeq 0.27$. Therefore, $\alpha_{\rm isob.}^2\approx 90$ per cent of the total variance is associated with isobaric,  $\alpha_{\rm adiab.}^2\approx 2$ per cent with adiabatic and $\alpha_{\rm isoth.}^2\approx 7$ per cent with isothermal perturbations on scales $\sim 30-80$ kpc, i.e. most of gas perturbation are associated with slow motions of gas in the inner region within the cluster core, and possibly with local changes of gas entropy. Even if the soft X-ray filament \citep[bright central feature towards the South, see][]{Fab01} is excluded from the images, the fluctuations are still found to be predominantly isobaric (dashed ellipse in Fig. \ref{fig:regr}).

Similar reasoning shows that, energetically, isothermal fluctuations appear to be dominant  ($\sim 88$ per cent of the total variance) on scales $\sim$ 30 kpc in the inner half-core region in the Centaurus cluster. On smaller scales, $\sim 10$ kpc, more than $50$ per cent of the total variance is associated with isobaric fluctuations. In both cases, the contribution of adiabatic fluctuations is $< 10$ per cent. Note that if we include the central, high-metallicity region in the analysis, the fluctuations become mostly isobaric on small and large scales (see Fig. \ref{fig:regr}).

Results for the Virgo and Perseus clusters are consistent with previous measurements by A16 and Z16 respectively. Namely, isobaric fluctuations are energetically dominant in Perseus, while fluctuations in Virgo are mostly isothermal (i.e. cavities filled with relativistic plasma). However, if the bright extended outflows in Virgo are included to the analysis, then the fluctuations become mostly isobaric.  

Poisson noise in the hard-band images limits our measurements of the power spectra on spatial scales smaller than $\sim 45$ kpc in A2029, A85 and Hydra A. For A2052 and A2199, the range of scales probed is narrow due to both high Poisson noise in the hard band and the effects of the underlying cluster model on large scales. Nevertheless, we repeat the analysis for these clusters as well. Fig. \ref{fig:reg_rest} shows a schematic R map with the loci of the measured $C$ and $R$ for the rest of clusters in the sample. The results for all clusters are also summarized in Table \ref{tab:cc}. In A2029, A85 and Hydra A, the gas perturbations are mostly isobaric, i.e. likely produced by slow gas displacements, while in A2199 fluctuations are predominantly isothermal, associated with bubbles and/or perturbations in the gravitational potential. The results for A2052 show a mixture of isobaric and isothermal fluctuations. In PKS0745-191,  systematic (the drop of spectrum on large scales) and statistical  (high Poisson noise on small scales) uncertainties on spectra in the hard band do not allow us to reach reliable conclusions about the nature of fluctuations. Even though shocks are found in some of these clusters, none  show predominantly adiabatic perturbations within the cool core. This is not very surprising, given that the shocks in our clusters are weak and do not fill the whole core. The situation could be different in some groups, such as NGC5813 \citep{Ran15}, elliptical galaxies and specific small region around the shocks in clusters (e.g. inner $30$ kpc regions in A2052 and Perseus).

So far, we have discussed the nature of fluctuations in the inner, half cool-core regions. In some clusters, the data allows us to probe the effective equation of state in the outer annulus, $r_{\rm cool}/2  < r < r_{\rm cool}$. Isobaric fluctuations appear to be dominant in A1795, A2052 and Perseus on spatial scales $\sim 60$--$90$, $\sim 46$--$86$ and $\sim 20$--$100$ kpc respectively, consistent with the sloshing of the gas. A85 shows mostly isothermal fluctuations on scales $\sim 90$ kpc and isobaric on scales $\sim 70$ kpc.

The results are based on $C$ and $R$ maps generated for the mean gas temperature within each considered region. If, instead, the maximum or minimum temperature is used, the dominant type of perturbations remains unchanged for the majority of clusters; however, the relative contribution of each type of perturbation may be different. Only in the inner regions of A2199 and A85 the maximal or minimal values of gas temperature change the dominant type of perturbations, from isothermal and isobaric to isobaric and isothermal, respectively.

\subsection{Velocity power spectra}
\label{sec:vel}

\begin{figure}
\includegraphics[trim=15 0 0 0,width=0.48\textwidth]{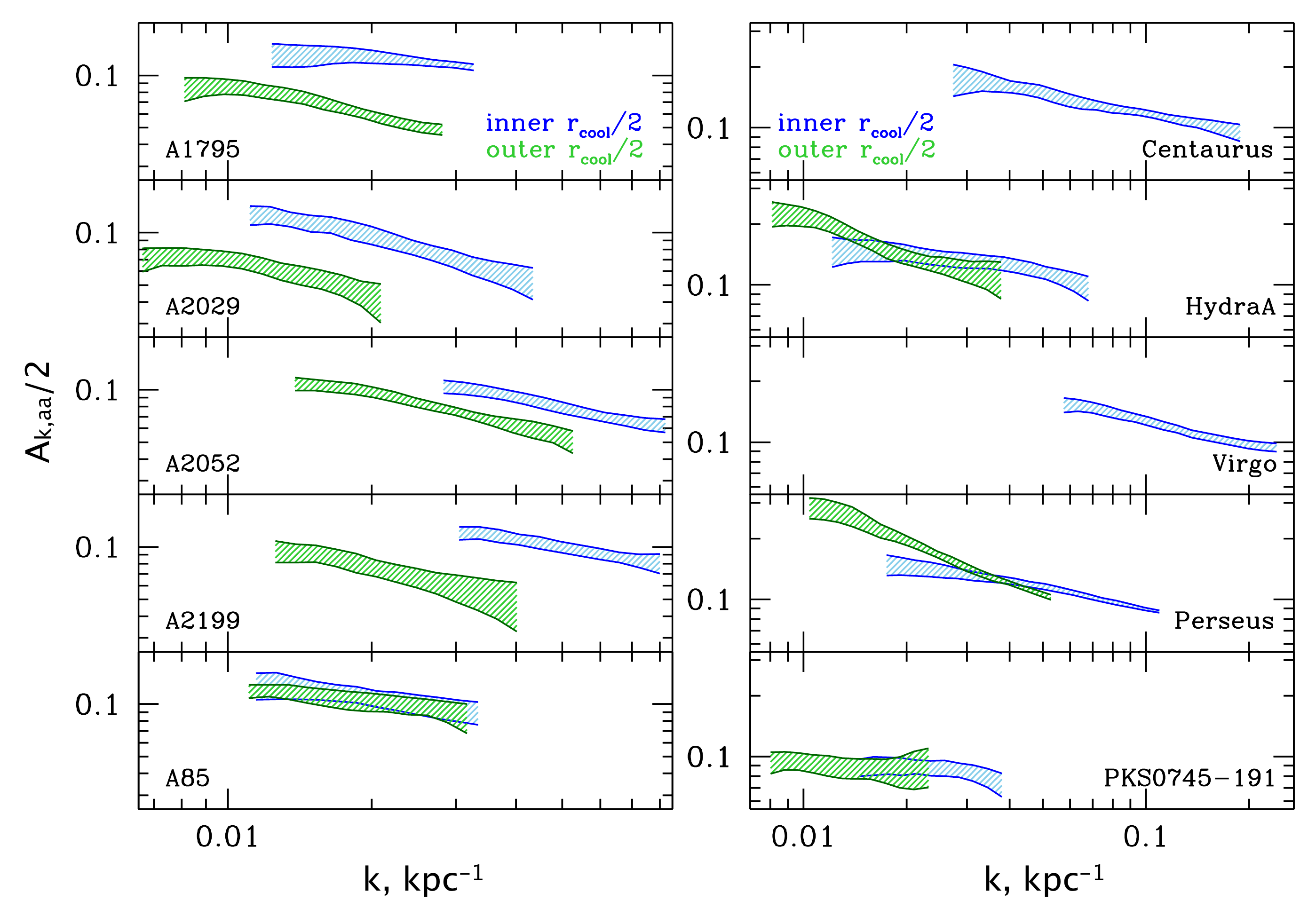}
\caption{The amplitude of density fluctuations as a function of a wavenumber measured in all clusters from our sample. Blue: inner, half-cool-core regions. Green: outer annuli. Only scales least affected by systematic uncertainties and Poisson noise are shown. 
\label{fig:amp_rest}
}
\end{figure}

\begin{figure*}
\includegraphics[trim=10 150 195 90,width=0.45\textwidth]{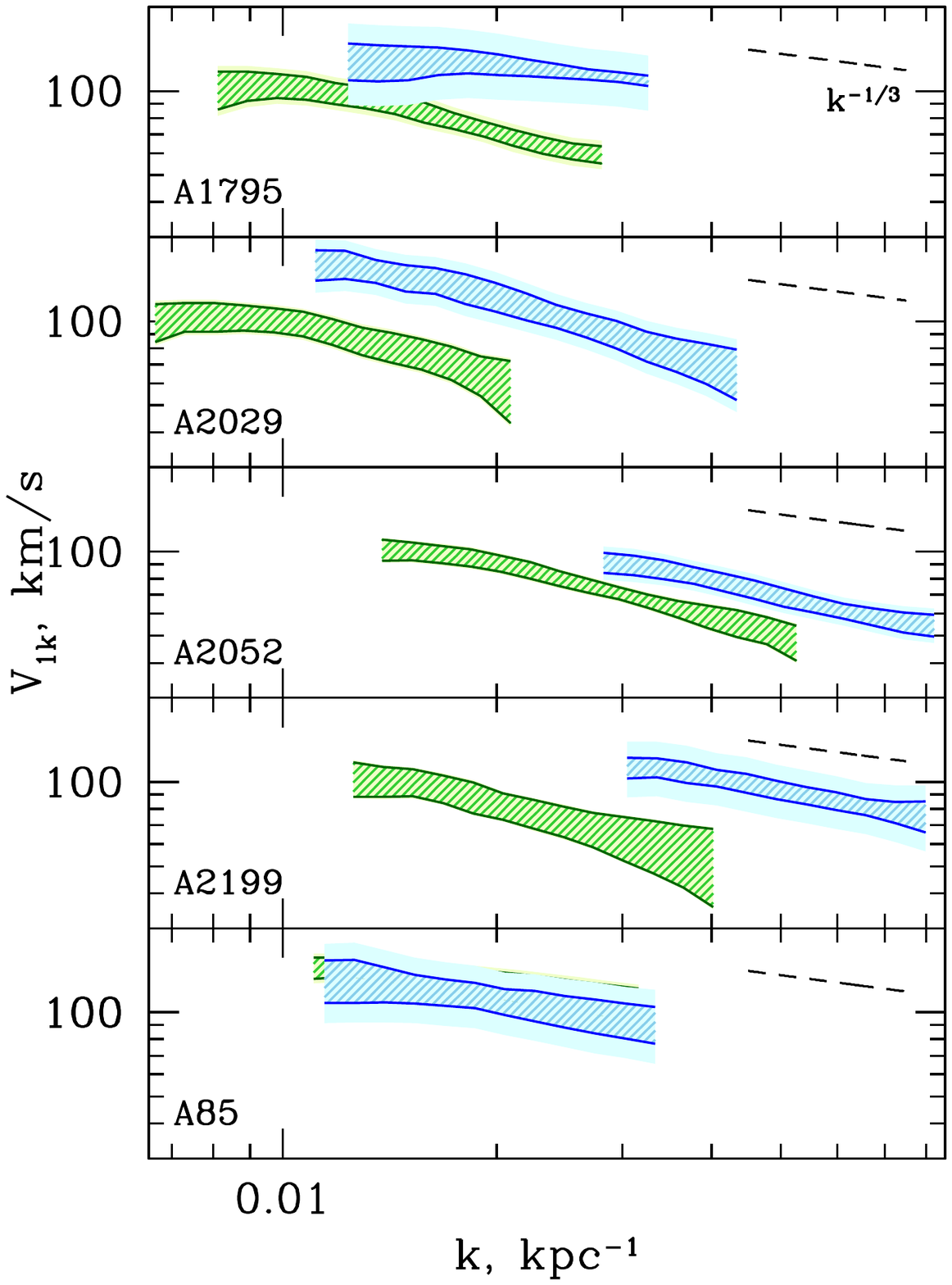}
\includegraphics[trim=10 150 195 90,width=0.45\textwidth]{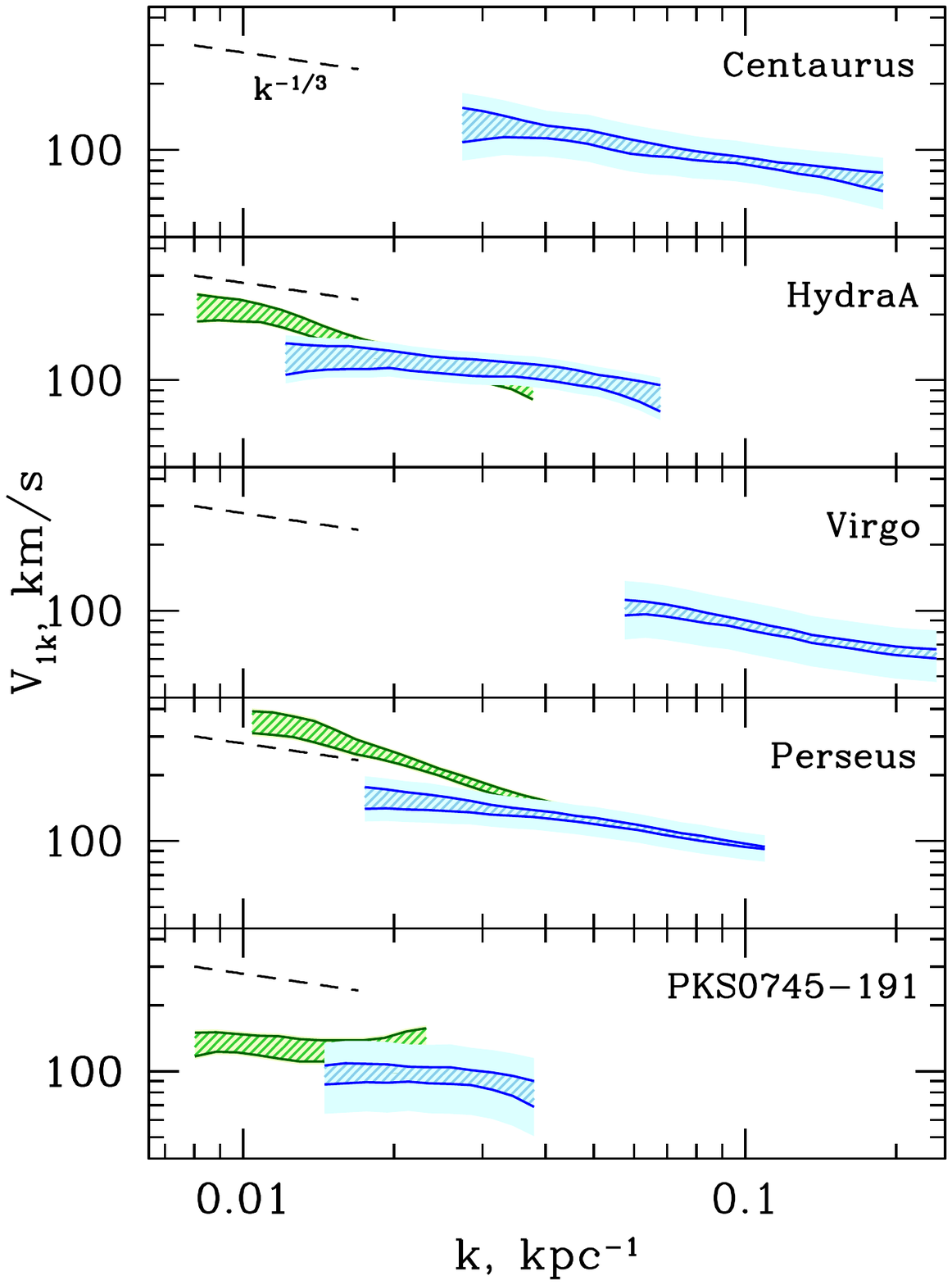}
\caption{Amplitude of one-component velocity of gas motions versus wavenumber $k=1/l$, measured for all clusters in our sample. Blue: velocities measured in the inner, half-cool-core regions. Green: same in the outer annulus of the cool core. Dashed lines show the slope of the amplitude for pure Kolmogorov turbulence (with arbitrary normalization). Hatched regions: velocities calculated using the mean sound speed in the gas. Solid regions: the spread of the velocity if maximal and minimal values of the sound speed are used. 
\label{fig:velocity}
}
\end{figure*}

The scale-dependent amplitude of density fluctuations, calculated in the inner and outer regions, is shown in Fig. \ref{fig:amp_rest}. In many clusters the amplitude of density fluctuations can be measured on spatial scales smaller than $\sim 15$ kpc, since Poisson noise in the soft-band images is significantly lower than in the hard-band images. The high-$k$ cutoff is determined by the PSF, and depends on the cluster redshift and offsets of the {\it Chandra} observations. We show only those wavenumbers where the PSF correction modifies the shape of the amplitude by less than $20$--$25$ per cent. The typical amplitude of density fluctuations is quite small, $<20$ per cent, even on relatively large spatial scales ($\sim 50$ kpc). It is  $<10-15$ per cent on scales of $\sim 10$ kpc. Such a small amplitude of density fluctuations, and their predominantly isobaric nature, indicate that the gas is perturbed gently. In the outer-core regions, the amplitudes are less than or comparable to the corresponding amplitudes in the inner regions on the same spatial scales (except for the amplitude on $\sim$ 60--70 kpc scale in Perseus and Hydra A). This could indicate the presence of several mechanisms that drive perturbations: central AGN in the inner region, and mergers and motions of substructure in outer one.

\begin{figure*}
\includegraphics[trim=0 0 0 0,width=\textwidth]{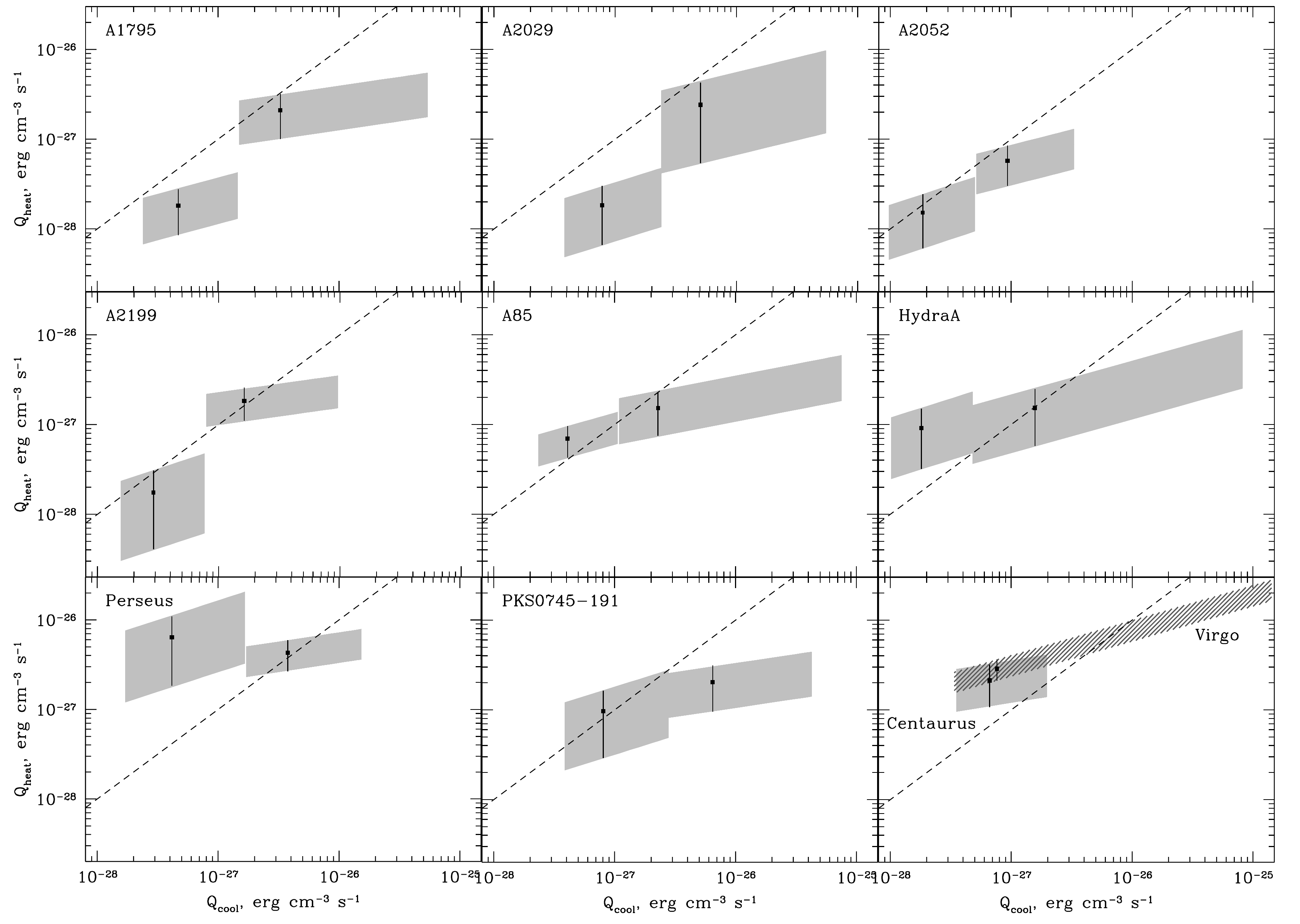}
\caption{Turbulent heating versus gas cooling rates in the cores of galaxy clusters from our sample. The size of each region reflects $1\sigma$ statistical and stochastic uncertainty in the heating rate, variations of density and temperature across each considered region within clusters and deviations of the measured spectral slope of the velocity spectra from the Kolmogorov law. For all clusters, except for Centaurus and Virgo, two measurements along the radius are shown. The dot in each region shows the volume-weighted average cooling within the region. Dashed lines correspond to the perfect balance between cooling and heating. Though for specific clusters the balance might be not perfect, on average, turbulent heating rate is comparable with radiative cooling rate.
\label{fig:ch}
}
\end{figure*}

In the previous section, we showed that in most clusters gas perturbations have an isobaric nature, i.e. are likely associated with slow displacements of gas in pressure equilibrium. Using the measured amplitude of density fluctuations (Fig. \ref{fig:amp_rest}) and the relation (4) between the amplitude of density fluctuations and one-component velocity of gas motions on each scale $1/k$, the amplitude (or power spectrum) of the velocity can be calculated; this is shown in Fig. \ref{fig:velocity}. Strictly speaking, the relation (4) was derived for slow gas motions in a stratified atmosphere (internal waves), which appear as isobaric perturbations in X-ray images.  However, in a few clusters our measurements suggest that the perturbations are predominantly isothermal or a mixture of isothermal and isobaric. Nevertheless, we keep these clusters in our sample, assuming that the scaling (4) is approximately correct. Indeed, ``isothermal" perturbations  could be due to bubbles of relativistic plasma inflated by an AGN. The potential energy associated with the bubbles follows approximately the same scaling with the observed variations of X-ray flux as the energy associated with internal waves (see A16 and Z16). As these bubbles rise through the cluster atmosphere, they can excite internal waves on scales comparable to the bubble size. Therefore, the scaling (4) may still approximately hold (in time/sample averaged sense), albeit with some scatter. 

Hatched regions in Fig. \ref{fig:velocity} show the results if the mean sound speed of the gas within the considered regions is used. Solid regions show the velocity spread when minimum and maximum values of the sound speed are used. Small-scale (e.g. less than 50 kpc) velocities of gas motions in the inner regions are quite low, $<100-150$ km/s, corresponding to 3D Mach numbers $M_k \lesssim 0.2-0.25$. On larger scales, 100 kpc and greater, the one-component velocity can be up to $200$--$300$ km/s ($M_k\sim 0.35$--$0.5$). The slopes of the spectra are consistent with that of Kolmogorov turbulence. However, the uncertainties on the measured spectra do not allow us to constrain the slopes precisely. The measured velocity amplitudes are consistent with the recent directly-measured velocities in the Perseus Cluster with the {\it Hitomi} satellite (see Section \ref{sec:hit}) and other measurements and upper limits \citep[e.g.][]{San10,San14,Zhu15,Wal15,Pin15,Hof16}.

\subsection{Turbulent heating}
\begin{figure}
\includegraphics[trim=30 150 30 80,width=0.45\textwidth]{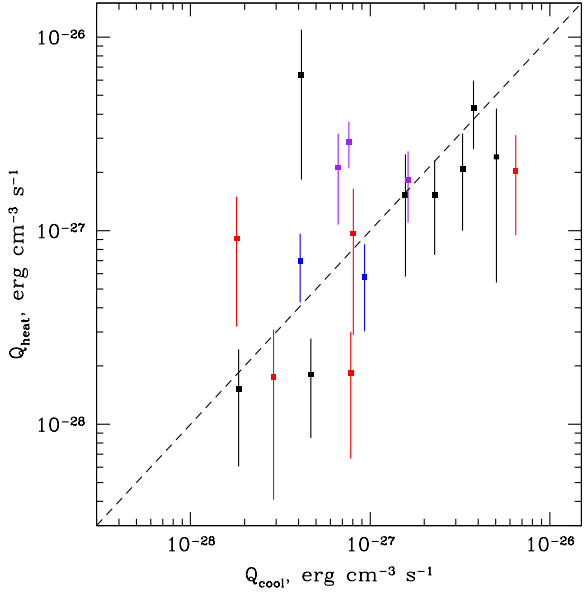}
\caption{Turbulent heating versus gas cooling rates for all clusters. Only volume-weighted values of the cooling rate are shown. Black/purple points indicate clusters, in which emissivity perturbations are predominantly of isobaric/isothermal nature; blue: mixture of isothermal and adiabatic fluctuations; red: unidentified.
\label{fig:ch_points}
}
\end{figure}

From the inferred velocity, we can estimate the heating rate of gas motions, which is $\propto \rho V^3/L$, where $\rho$ is the gas mass density, $V$ is the characteristic velocity and $L$ is the energy-containing scale. Since several characteristic scales are usually present in cluster atmospheres, the energy-containing scale is not clearly defined. Therefore, we calculate the heating rate assuming purely hydrodynamic volume-filling turbulence and following the Kolmogorov scaling. Namely, $Q_{\rm turb}=C_Q \rho V_{1,k}^3 k$, where $C_Q=3^{3/2}2\pi/(2C_K)^{3/2}\approx 5$ is a dimensionless constant related to the Kolmogorov constant $C_K$ that accounts for our conventions and $\rho$ is the the gas density \citep[see][for details]{Zhu14b}. We compare the turbulent heating rate with the cooling rate, defined as $Q_{\rm cool}=n_en_i\Lambda(T)$, within the cool cores of the clusters in our sample in Fig. \ref{fig:ch}. Since in the inner regions the gas density profile is often very steep towards the center, variations of $Q_{\rm cool}$ are quite large.  Black points show the results if volume-weighted gas density and temperature are used for the calculations of the cooling rate. Fig. \ref{fig:ch_points} shows all the measurements plotted together. Although the scatter and uncertainties are large, on average, there appears to be an approximate balance between $Q_{\rm turb}$ and $Q_{\rm cool}$, suggesting that the dissipation of gas motions provide enough heat to compensate for gas cooling losses, or at least significantly contribute to energy balance. These results are in line with earlier conclusions based on the detailed analyses of Perseus, Virgo and Centaurus clusters \citep{Zhu14b, Wal15}.   

There are many underlying assumptions in the above calculations. First, we neglect gas viscosity. To check this assumption, we calculate the dissipative (Kolmogorov) scale $l_{\rm K}=\nu_{\rm kin}^{3/4}/(Q_{\rm cool}/\rho)^{1/4}$, where $\nu_{\rm kin}$ is the kinematic viscosity calculated for unmagnetized gas, assuming the cooling - heating balance. In all clusters, the dissipative scale is $<10$ kpc (typically $\sim$ few kpc) in the inner regions and $<20$ kpc in the outer regions, i.e. significantly less than the smallest scales that we probe. If the plasma is magnetized, then the viscous scale is predicted to be even smaller, as various plasma instabilities will sustain perturbations down to gyroscales \citep[e.g.][]{Sch02,Sch06}. 

Another assumption is that the injection scale of gas motions is larger than, or at least comparable to, the scale at which the eddy turnover timescale becomes smaller than the buoyancy timescale. Assuming approximate balance between cooling and turbulent heating rates, we estimate the Ozmidov scale as $N^{-3/2}(Q_{\rm cool}/\rho)^{1/2}$, where $N$ is the Brunt-V$\rm \ddot a$is$\rm \ddot a$l$\rm \ddot a$ frequency in the cluster atmosphere, which depends on the acceleration of gravity and the entropy scale height. Given the thermodynamic profiles of clusters, we find that, typically, the Ozmidov scale is smaller than $10$--$20$ kpc in the inner-core regions, and up to $30$ kpc in the outer regions. Both are within the span of scales that we probe in each cluster.

\section{Discussion}
\subsection{Energy content of emissivity fluctuations and dissipation timescale}
Assuming that the total variance of fluctuations is the sum of fluctuations produced by bubbles, weak shocks/sound waves and slow motions of gas in stratified atmosphere (internal gravity waves), we can write the total energy associated with gas perturbations relative to the thermal energy of gas as $E_{\rm per}/E_{\rm th}=(E_{\rm b}+E_{\rm sw}+E_{\rm iw})/E_{\rm th}$. As shown by A16 and Z16, the ratio of energy in bubbles and thermal energy is related to the total variance of the gas density fluctuations, $(\delta \rho/\rho)_{\rm b}$, as
\be
\frac{E_{\rm b}}{E_{\rm th}}=\frac{\gamma _{\rm b}(\gamma -1)}{\gamma_{\rm b}-1}\left\langle\left(\frac{\delta\rho}{\rho}\right)_{\rm b}^2\right\rangle,
\label{rel:ebub}
\ee
where $\gamma_{\rm b}=4/3$ is the  adiabatic index of the hot relativistic gas inside the bubble and $\gamma=5/3$ is the adiabatic index of the ambient gas. For sound waves, the corresponding relation is
\be
\frac{E_{\rm sw}}{E_{\rm th}}=\gamma(\gamma-1)\left\langle\left(\frac{\delta\rho}{\rho}\right)_{\rm sw}^2\right\rangle,
\label{rel:esw}
\ee 
while for internal waves it becomes
\be
\frac{E_{\rm iw}}{E_{\rm th}}=\gamma\left\langle\left(\frac{\delta\rho}{\rho}\right)_{\rm iw}^2\right\rangle.
\label{rel:egw}
\ee 
Relations (\ref{rel:ebub}--\ref{rel:egw}) have the form $E_{\rm i}/E_{\rm th}\propto C \left\langle\left(\frac{\delta\rho}{\rho}\right)_{\rm i}^2\right\rangle$, where the constant $C$ is of the order of few units. 
Since fluctuations are mostly isobaric (or isobaric/isothermal) we use the relation (\ref{rel:egw}) to estimate $E_{\rm per}/E_{\rm th}$. The total variance of density fluctuations can be calculated by integrating the measured power spectrum of emissivity fluctuations in the soft band, $P_{k,aa}$, 
\be
\left\langle\left(\frac{\delta\rho}{\rho}\right)^2\right\rangle=\frac{1}{2}\int\limits_{k_{\rm min}}^{k_{\rm max}} P_{k,aa}4 \pi k^2 dk.
\ee
$P_{k,aa}$ divided by 2 is shown in Fig. \ref{fig:amp_rest} in the units of scale-dependent amplitude, as are $k_{\rm min}$ and $k_{\rm max}$ for each cluster. For most clusters in our sample, the total variance of density fluctuations is $\langle(\delta \rho / \rho)^2 \rangle \sim 0.01$--$0.03$, which implies that the energy in perturbations (non-thermal energy) is $\sim 2$--$5$ per cent of the thermal energy. This agrees with other previous measurements in cluster cores (e.g. Hofmann et al. 2016, A16, Z16, Hitomi Collaboration et al. 2016). Interestingly, the results are also consistent with the contribution of non-thermal pressure measured in small groups and elliptical galaxies \citep{Chu08,Wer09,deP12,Ogo17}. In the outer regions in Perseus and Hydra A and in the inner region in Centaurus, the energy in perturbations is slightly higher, $\sim 7$--$12$ per cent of the thermal energy. Given that the integration range is limited, these estimates could be lower limits.  

\begin{figure*}
\includegraphics[trim=0 0 0 0,width=0.95\textwidth]{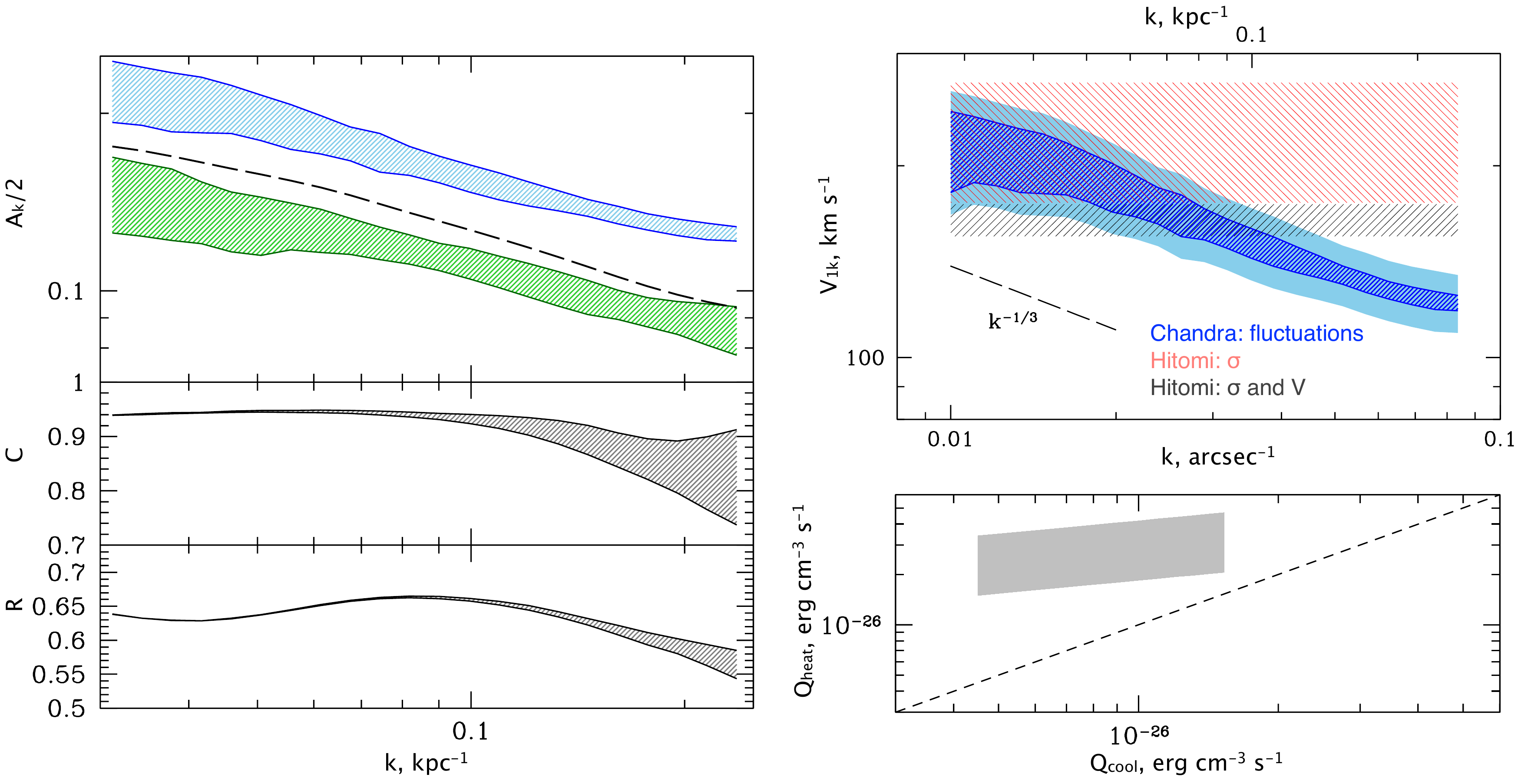}
\caption{The analysis of gas perturbations in the Perseus Cluster in the region observed by the {\it Hitomi} satellite \citep[outer region; see Fig. 3 of][]{Hit16}. {\bf Left:} results of the auto- and cross-spectra analysis, using the same notation and color scheme as Fig. \ref{fig:cc}. {\bf Right, top:} one-component, scale-dependent velocity amplitude measured from the analysis of gas fluctuations (light and dark blue), see Fig. \ref{fig:velocity} for notations. The {\it Hitomi} measurements are plotted with hatched regions (only amplitudes, no information on spatial scales): the line-of-sight velocity dispersion in black, and the total velocity estimated through line broadening and variation of line centroid in red. {\bf Right, bottom:} cooling rate vs turbulent heating rate measured from the velocity spectrum in the top panel. See Section \ref{sec:hit} for details.
\label{fig:hit}
}
\end{figure*}

Given an approximate balance between the radiative cooling and turbulent dissipation rates (see Fig. \ref{fig:ch_points}), one can estimate the dissipation timescale needed to convert the energy in perturbations into heat, namely $t_{\rm diss}=t_{\rm cool}\times E_{\rm per}/E_{\rm th}$. $E_{\rm per}/E_{\rm th}$ varies between $2$--$12$ per cent, therefore, $t_{\rm diss}$ is $\sim 0.04$--$0.4$ Gyrs given that the typical cooling time is a few Gyrs.

In the above estimates we neglect thermal conduction, which could be particularly important for the most massive clusters in our sample, and gas viscosity \citep[e.g.][]{Rus10,ZuH13,Gas14}. We also do not take into account radiative cooling, which can produce low-entropy structures that appear as isobaric perturbations. The inclusion of these effects requires additional theoretical studies and is beyond the scope of this paper.

\subsection{Comparison of the velocity measurements with the {\it Hitomi} results}
\label{sec:hit}

Recent {\it Hitomi} observations have shown that the gas in the core of Perseus has a line-of-sight velocity dispersion (non-thermal line broadening) of $\sigma=164\pm10$ km/s and a gradient in the line-of-sight velocity (variations of line centroids) of $V=150\pm70$ km/s in the region $\sim 30$--$60$ kpc from the cluster center \citep{Hit16}. It is interesting to compare this result with our indirect velocity measurements. The left panel of Fig. \ref{fig:hit} shows the results of fluctuation analysis in Perseus in the outer region observed by {\it Hitomi} \citep[see Fig. 3 of][]{Hit16}. The amplitude of fluctuations in the soft band is larger than in the hard band on scales between $\sim 5$--$30$ kpc, suggesting that fluctuations are mostly isobaric. The locus of the measured coherence and ratio in the $C-R$ maps reveals that $\sim 81$ per cent of the total variance is indeed associated with isobaric perturbations, and $\sim 12$ and $\sim 7$ per cent with adiabatic and isothermal fluctuations, respectively. This implies that the ratio of non-thermal energy (energy in perturbations) to the thermal energy is $\sim 9$ per cent, with a scatter of $\sim 2$ per cent. This is consistent with the {\it Hitomi} results, namely the measured ratio of turbulent to thermal energy of $\sim 8$ per cent (if both velocity dispersion and large-scale motions are taken into account). 

The right panel of Fig. \ref{fig:hit} shows the one-component, scale-dependent velocity amplitude, corrected for the $\it Chandra$ PSF on small scales. The  mean one-component velocity is $\sim 215$ km/s on $\sim 30$ kpc scales, and $\sim 120$ km/s on $\sim 5$ kpc scales with some scatter. On the same plot we show the velocity dispersion, $\sigma$, measured with {\it Hitomi}, as well as the total velocity, estimated as $\sqrt{\sigma^2+V^2}=222\pm48$ km/s. Direct comparison of the $\it Chandra$ and $\it Hitomi$ velocities is not straightforward, since we do not know what scales contribute most to the $\it Hitomi$ measurements. However, if the dominant scale is between $\sim$ 10 and several 10s of kpc, a reasonable assumption given the size of the observed region, its distance from the center, and the size of the largest structures (bubbles), then the results from both methods agree. The energy associated with such gas motions, once dissipated, is more than enough to offset radiative cooling (see the bottom-right panel of Fig. \ref{fig:hit}). 

If the dominant scale contributing to the Hitomi measurement is comparable to the size of the cool core ($\sim 150$ kpc or more), and the slope of the velocity power spectrum is the same on large spatial scales, then turbulent heating cannot balance the cooling. This scenario is unlikely, however, as the projected linear size of the considered region is less than $30$ kpc and the effective length along the line of sight, which provides the dominant contribution to the line flux, is smaller than $100$ kpc at distances of $30$--$60$ kpc from the center \citep[][]{Zhu12}. If, instead, the dominant scale is smaller than $\sim 7$ kpc, the measured level of turbulence overheats the gas, i.e. the heating rate in this case is greater than the cooling rate. To summarize, the fluctuation analysis provides velocity measurements, which are consistent (within the uncertainties, which could be up to a factor of 2) with the direct measurements.
 
\section{Conclusions}
In this paper, we measured the effective equation of state of gas perturbations, the velocity power spectra of gas motions and discussed the role of turbulent dissipation in the heating of gas in the brightest core regions of nearby galaxy clusters. The analysis is based on the measurements of power spectra of X-ray surface brightness and density fluctuations. Our sample includes ten clusters that have {\it Chandra} data deep enough to probe fluctuations on spatial scales between $\sim 10$ and few tens of kpc,  and that show observational indications of radio-mechanical AGN feedback. We carefully treated uncertainties crucially important for the measurements on such small spatial scales, which are associated with Poisson noise, unresolved point sources, the {\it Chandra} PSF, and projection effects. Our main conclusions are summarized below.

(i) In all clusters gas perturbations in the inner half cool core regions are predominantly isobaric or isothermal on spatial scales between $\sim 10$ and $60$ kpc, i.e. likely associated with slow gas motions and bubbles of relativistic plasma (X-ray cavities). Adiabatic perturbations associated with weak shocks constitute less than 10 per cent of the total variance. At least in half of the considered clusters, the gas is, undoubtedly, disturbed by the central AGN activity. Subdominant contribution of adiabatic fluctuations in these clusters supports a model of a gentle AGN feedback rather than the more explosive scenarios that have been explored by some numerical simulations. In the outer parts of cool cores, fluctuations are mostly isobaric. If perturbations in these regions are driven by mergers and motions of galaxies instead of the central AGN activity, the results are consistent with the nature of  perturbations seen in cosmological numerical simulations of galaxy clusters \citep{Zhu13}.  

(ii) The energy in perturbations (non-thermal energy) is measured to be $\sim$ 5 per cent of the thermal energy of the hot gas in the inner, half cool core regions. In the outer core regions, where mergers may substantially contribute, it can be up to $12$ per cent. These results are consistent with previous measurements, including the {\it Hitomi} results \citep{Hit16}.

(iii) The typical amplitude of density fluctuations is small, less than $\sim 20$ per cent on scales $\sim 50$ kpc. On scales of $\sim 10$ kpc, the amplitude is typically $\sim 10$ per cent and less. 

(iv) Typical one-component velocity of gas motions measured from the density power spectra is $\lesssim 100-150$ km/s on scales smaller than $\sim 50$ kpc. On larger scales of $\sim 100$ kpc, the velocities can be up to $\sim 300$ km/s. Within the uncertainties, the results agree with the direct velocity measurements from the {\it Hitomi} observations.

(v) Regardless of the source that drives gas motions, the dissipation of their energy provides enough heat to compensate gas cooling losses. The result is valid on average; for specific clusters/regions the balance can be not perfect. The dissipation timescale of gas perturbations is $\sim 0.04$--$0.4$ Gyrs.    

(vi) Our results are consistent with previous observational results and support numerical models in which the energy from central AGN is transported slowly to the ICM through the inflation of bubbles of relativistic plasma. 

The full power of the techniques discussed here will be revealed in future with the next generation X-ray observatories like {\it Athena} and {\it Lynx} that will have effective areas at least an order of magnitude larger than the current X-ray observatories. The high spatial resolution of the ${\it Lynx}$ observatory and its stable PSF across the field of view will open a possibility to probe microphysical scales in objects of different masses at different redshifts, allowing detailed studies of transport processes, MHD, and plasma effects. 

\section{Acknowledgements}
We acknowledge support from NASA through Chandra Award Number AR7-18015X, issued by the Chandra X-ray Observatory Center, which is operated by the Smithsonian Astrophysical Observatory for and on behalf of NASA under contract NAS8-03060. IZ thanks E. Churazov for providing constructive comments on the manuscript.

\appendix

\label{lastpage}  
\end{document}